\documentclass{ws-procs9x6}
\usepackage{graphicx}

\newcommand{\be}{\begin{eqnarray}}
\newcommand{\ee}{\end{eqnarray}}

\newcommand{\ra}{\rangle}

\begin{document}

\title{Linking Chiral and 
Topological Aspects of the Quark-Quark Interaction, in 
QCD\footnote{\uppercase{T}alk given at the 
\uppercase{I}nternational \uppercase{C}onference on \uppercase{C}olor
\uppercase{C}onfinement and 
\uppercase{H}adrons in  
\uppercase{Q}uantum \uppercase{C}hromodynamics
 (\uppercase{C}onfinement 2003), \uppercase{T}okyo, 
\uppercase{J}uly 2003.}}

\author{P. Faccioli\footnote{Speaker}}
\address{E.C.T.* 
Strada delle Tabarelle 286,Villazzano (Trento), I-38050, Italy.\\
E-mail: faccioli@ect.it}
\author{T.A. DeGrand}
\address{University of  Colorado, Boulder, CO 80309, USA.\\
E-mail: degrand@aurinko.colorado.edu}

\maketitle
\abstracts{We present the results of a recent investigation of
the short-distance structure of non-perturbative
interaction between light quarks. We analyze the
connection between topology and the dynamics 
responsible  for the spontaneous breaking of chiral symmetry.
By measuring some appropriate correlation functions on the lattice,
we provide compelling evidence for instanton-induced dynamics.}

\section{Introduction}
Understanding the light-quark dynamics at all scales is a fundamental and 
open problem, in high-energy physics. 
In particular, the improving our knowledge of the non-perturbative
interaction at intermediate and short distances has become essential, 
in order to explain the results of
recent experiments on pion and proton electro-magnetic formfactors,
performed at Jefferson Laboratory \cite{JLAB}.
Indeed, these high-precision measurements have shown that the 
perturbative  regime in QCD does not set-in until very large 
values of the momentum transfer ($Q^2>6~GeV^2$). 
The delay of the onset of the asymptotic regime implies that there are 
strong, short-scale non-perturbative interactions, inside ordinary hadrons.

It is commonly accepted 
that the physics of the light-quark sector of QCD 
is strongly influenced by the dynamics 
responsible for the  chiral symmetry breaking (CSB).
The characteristic energy scale associated to such a non-perturbative 
phenomenon is 
$4\,\pi f_\pi \sim 1.2~GeV$, 
sizably larger than the typical confinement scale, $\Lambda_{QCD}$.
Such a separation of scales justifies attempting to understand the 
short-distance non-perturbative structure of light-hadrons, 
without having simultaneously to understand microscopic origin of color 
confinement. 
On the other hand, from the observation that $4\,\pi\,f_\pi \sim m_{\eta'}$
it follows that topological effects should be included in any effective 
description of the light-quark dynamics.

An effective interaction\footnote{For 
simplicity, in this work we shall restrict
to the $N_f=2$ case.}, which can simultaneously solve the U(1) problem and
explain the CSB was derived by 't Hooft \cite{'tHooftU1},
by computing  the Euclidean QCD generating functional 
in the semi-classical limit  
(i.e. accounting for small perturbations around the instanton solution):
$L_{'t~H}~=~G_{\bar{\rho},\bar{n}}\,
\left([\,(\psi^\dagger\,\tau^-_a\,\psi)^2 
- (\psi^\dagger\,i\gamma_5\,\tau^-\,\psi)^2\, ]
+\frac{1}{2(2\,N_c-1)}\,(\psi^\dagger
\,\sigma_{\mu\,\nu}\,\tau_a^-\,\psi)\right),$
where $\tau^{-}=(\vec{\tau},i)$ ($\vec{\tau}$  are isospin Pauli matrices), and
$G_{\bar{n}\,\bar{\rho}}$ is a coupling constant depending on
the typical density of instantons in the vacuum $\bar{n}$,
and  their typical size $\bar{\rho}$. Notice that the 
finite size of the instanton field
provides a natural cut-off scale for the interaction.

Unfortunately, the instanton density and size
cannot be computed systematically in QCD, so the
semi-classical 't Hooft interaction  is not completely specified. 
The Instanton Liquid Model (ILM)\cite{shuryakrev} 
consists of assuming that the vacuum is saturated
by instantons with typical size $\bar{\rho}\simeq 1/3$~fm and density 
$\bar{n}\simeq 1\,\textrm{fm}^{-4}$.

In a number of recent works
it was shown that instantons can quantitatively explain the observed
deviation of the pion and nucleon electromagnetic formfactor from the
perturbative prediction, at high momentum 
transfer\cite{formfactors}.
In the present work, we complement that analysis, and
use lattice simulations to look directly  for the specific
signatures of the 't~Hooft interaction.

\section{Instantons and Chirality-Mixing Interactions}
\label{Dirac}
A characteristic property of the 't Hooft effective vertex is that it 
mixes only quarks of {\it different chirality}.
Hence, if instantons are the dominant sources of the 
non-perturbative interaction, then one should be able to observe that
quarks change their chirality after each non-perturbative interaction.
On the contrary, if the non-perturbative vertex has the same structure
of the perturbative gluon exchange (i.e. a purely vector or axial-vector 
coupling), then a single interaction should not flip the chirality of quarks.
In this case, chirality mixing should arise only as a consequence of  
mass  generation, i.e. from a genuine many-body effect.

Based on this observations, one is led to consider the ratio:
\be
\label{RNS}
R^{NS}(\tau):=\frac{A^{NS}_{flip}(\tau)}{A^{NS}_{non-flip}(\tau)}=
\frac{\Pi_\pi(\tau)-\Pi_\delta(\tau)}
{\Pi_\pi(\tau)+\Pi_\delta(\tau)},
\ee
where $\Pi_\pi(\tau)$ and $\Pi_\delta(\tau)$ are pseudoscalar (PS) and scalar 
(S) iso-triplet
two-point correlators,
related to the currents $J_\pi(\tau):=\bar{u}(\tau)
\,i\gamma_5\,d(\tau)$ 
and $J_\delta(\tau):=\bar{u}(\tau)\,d(\tau)$.
If the propagation is chosen along the  
time direction, $A^{NS}_{flip(non-flip)}(\tau)$ represents 
the probability amplitude for a 
$|q\,\bar{q}\ra$ pair with isospin 1 to be found after a 
``time'' interval $\tau$ in a state
in which the chirality of the quark and anti-quark 
is interchanged (not interchanged).
Notice that the ratio $R^{NS}(\tau)$ 
must vanish as $\tau\to 0$ (no chirality flips),
 and must approach 1 as  
$\tau\to \infty$ (infinitely many chirality flips).

 It can be shown  that the correlator (\ref{RNS}) 
 receives no perturbative contribution.
Moreover, a model-independent spectral  analysis 
indicated that the interaction responsible for dynamical chirality-mixing
is mediated by topological fields, and that the 
rate of chirality flips in a quark-antiquark system
is proportional to the mass of the $\eta'$ meson\cite{chimix}.

\begin{figure}
\includegraphics[scale=0.2,clip=]{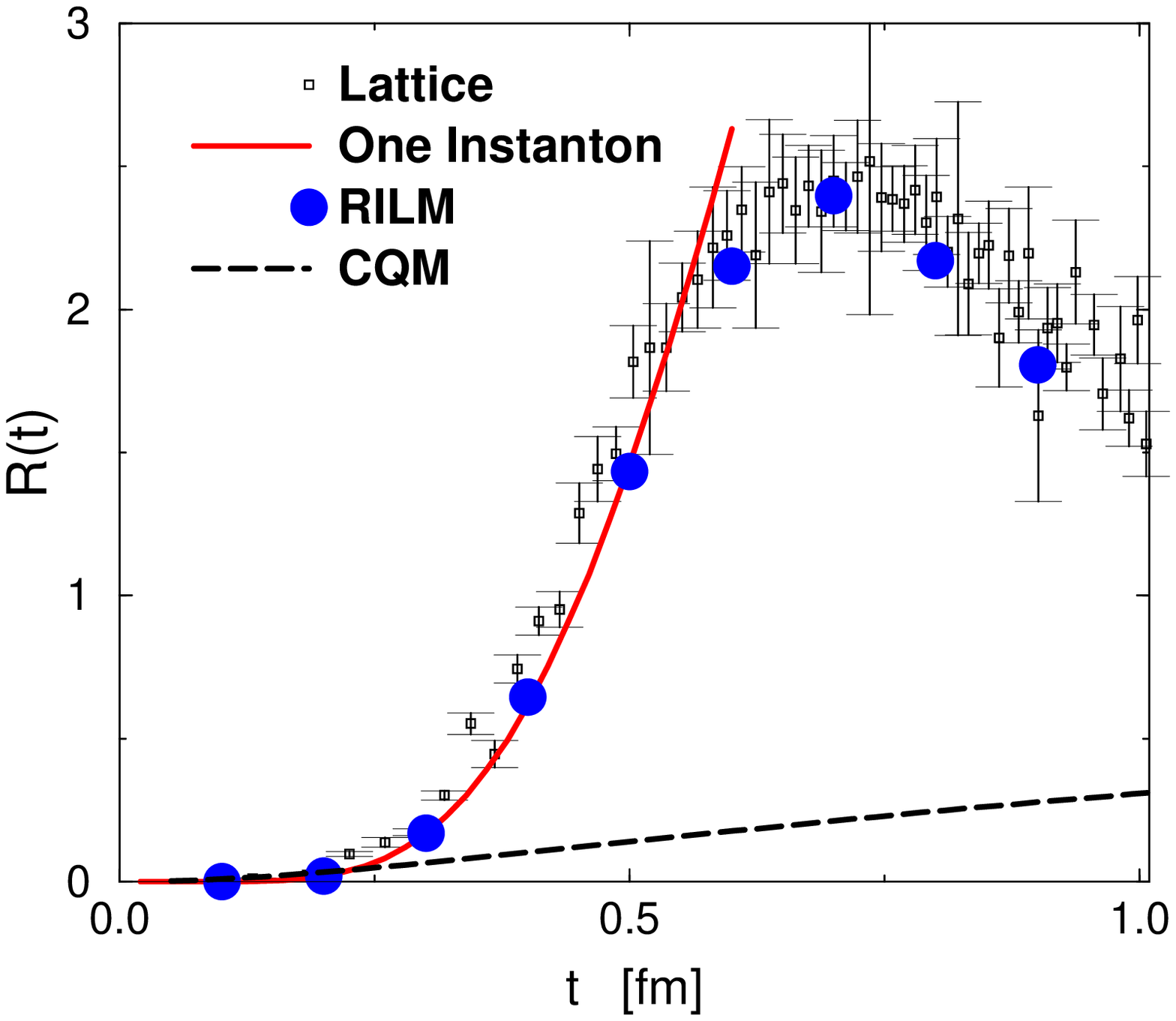}
\hspace{1.5cm}
\includegraphics[scale=0.2,clip=]{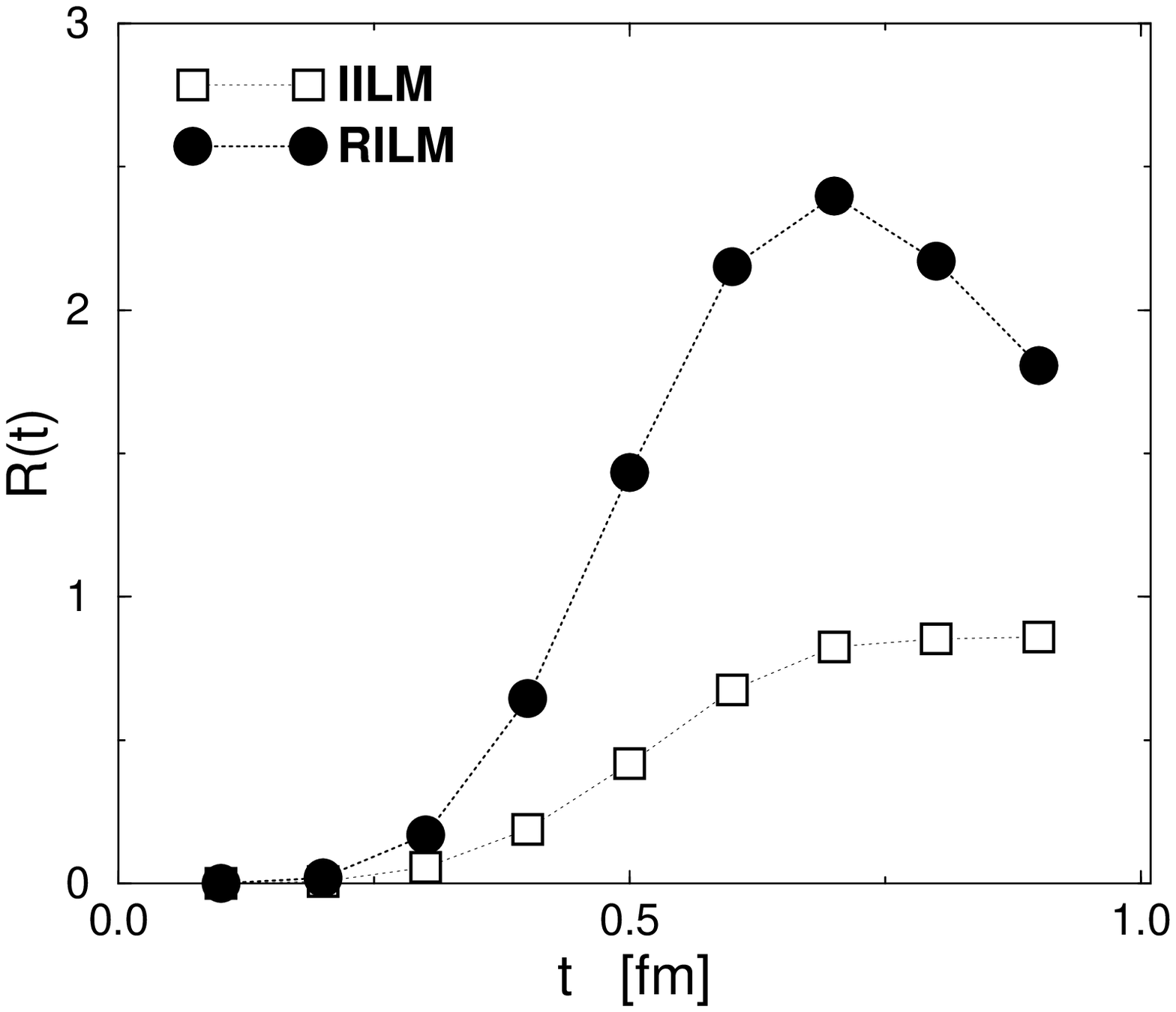}\\
(A)\hspace*{4.5cm} (B)\\
\caption{(A): The chirality-flip correlator 
in lattice QCD (square points) and in the RILM (circles).
The solid line represent the single-instanton contribution. 
The dashed curve was obtained from  two free ``constituent''
quarks with a mass of
400 MeV.
(B): Suppression of chirality flipping events, due to fermion-loop 
exchange in the ILM. Circles are RILM (quenched) results, squares are IILM 
(unquenched) results.}
\label{mainfig}
\end{figure}

The  chirality flip ratio $R^{NS}(\tau)$ was recently investigated in QCD, 
by means of lattice simulations performed with chiral fermions \cite{scalar}.
The results are the square points plotted in Fig.~\ref{mainfig}(A). 
As expected, the lattice data interpolate between 0 and 1. 
We observe that after few fractions of a fermi 
the quarks are more likely to be found
in a configuration in which their chiralities is flipped, than to be
found in their initial configuration.
The presence of a maximum in $R^{NS}(\tau)$, 
and the subsequent fall-off toward 1 
have a very simple explanation in the ILM: 
if quarks propagate in the vacuum for a time comparable with
the typical distance between 
two neighbor instantons (i.e. two consecutive 't Hooft interactions), 
they have a large probability of crossing the field of the closest
pseudo-particle. 
If so happens, they must necessarily flip their chirality.
So, after some time, the quarks are most likely to be found in the
configuration in which their chirality is flipped.
On the other hand, 
if one waits for a time much longer than 1~fm, then the quarks will ``bump''
 into many other such pseudo-particles,
experiencing several more chirality flips. Eventually, either chirality 
configurations will become equally probable 
and $R^{NS}(\tau)$ will approach 1.

Let us now consider the result for  $R^{NS}(\tau)$ obtained 
in the Random Instanton Liquid Model\footnote{In such a version of the ILM
quarks are assumed to propagate in a vacuum populated by an ensemble of
randomly distributed instanton and anti-instantons. 
It can been shown that the RILM accounts for the 't Hooft interaction to
all orders, but neglects quark loops (quenched approximation).}
(circles in Fig.~\ref{mainfig}(A)). 
The agreement with the lattice results is impressive, implying not only that
instantons are the relevant sources of short-range 
chirality-mixing interaction, but also
that the phenomenological estimates $\bar{\rho}\simeq 1/3$~fm, 
$\bar{n}\simeq~1\,\textrm{fm}^{-4}$ are realistic.

\section{Instanton, Unitarity and Quark Loops}
\label{loops}

In this section we discuss an interesting link between topology 
and unitarity, which emerges when one 
analyzes the contribution of quark loops to chirality flips.
We recall that lattice 
results reported in section \ref{Dirac} 
have been obtained in the quenched approximation. 
Unitarity requires that all correlation functions must be positive definite
and this implies  that $R^{NS}(\tau)\le~1$, at all times.
The existence of a maximum in which the chirality flip ratio 
is greater than one is a reflection of the fact that,
in the quenched approximation,  the unitarity of the theory is lost.
We necessarily conclude that the fermionic determinant suppresses some
chirality-flipping events, which are otherwise allowed in the
quenched approximation.
Clearly, {\it realistic descriptions of the effective quark-quark interaction
 must reproduce a dramatic enhancement of the 
chirality flipping amplitude, when quark loops are suppressed.}

Quite remarkably, 
this phenomenon is naturally explained in the instanton picture.  
If quark loops are allowed, then
instantons and anti-instantons can interact through
fermion exchange.
Such an interaction generates an attraction between 
instantons and anti-instantons
leading to a screening of the topological charge.
As a result of such a screening, quarks crossing the field of an instanton 
are very likely to find, in the immediate vicinity, an
anti-instanton which restores their initial chirality 
configuration.
In Fig.~\ref{mainfig}(B) we compare the chirality flipping ratio
 $R^{NS}(\tau)$ obtained from a quenched (RILM)
and unquenched (IILM\footnote{Interacting Instanton Liquid Model}) 
version of the ILM.
We observe that, with the inclusion of the fermionic determinant, 
the unitarity condition $R^{NS}(\tau)<1$ is restored. 
We  stress that, although such a restoration  
must necessarily take place in QCD, it represents a remarkable success 
of the ILM, which is not a unitary field theory.

\section{Conclusions}
\label{conclu}

We have presented a study of the quark 
dynamics, at intermediate and short distances, based
on the results of lattice QCD simulations with chiral fermions.
Our analysis provides strong evidence that instantons represent the main source
of non-perturbative interaction between light quarks.
We studied the contribution of quark loops,
by comparing the results of quenched and full simulations. 
We observed  dramatic quenching effects that can be  
naturally explained by instantons 
and suggests an interesting link between chiral dynamics, 
unitarity and topology.


\begin{thebibliography}{99}


\bibitem{JLAB} J. Volmer {\it et al.} [The Jefferson Laboratory $F_\pi$
Collaboration], 
Phys. Rev. Lett. \textbf{86} (2001) 1713. O. Gayou, \emph{et al.}, 
Phys. Rev. Lett.
\textbf{88} (2002) 092301.
\bibitem{'tHooftU1}     
G.'t Hooft,    Phys. Rev.,  \textbf{D14}(1976), 3432.
\bibitem{shuryakrev}
T. Schaefer and E.V. Shuryak, Rev. Mod. Phys. \textbf{70} (1998) 323.
[arXiv:hep-ph/9610451].
\bibitem{formfactors}
P. Faccioli, A. Schwenk and E.V. Shuryak 
[arXiv:hep-ph/0309129].
P. Faccioli, A. Schwenk and E.V. Shuryak, Phys. Rev.
\textbf{D67} (2003) 113009  
[arXiv:hep-ph/0202027].
P. Faccioli, A. Schwenk and E.V. Shuryak, Phys. Lett. \textbf{B528} (2002) 34
[arXiv:hep-ph/0205307].
\bibitem{chimix}
P.~Faccioli,
[arXiv:hep-ph/0211383].
%
\bibitem{scalar} P. Faccioli and T. A. DeGrand, 
[arXiv:hep-ph/0304219], to appear on Phys. Rev. Lett.
%
%
%
\bibitem{Martinez} H. Martinez, Ph.D. Thesis, M.I.T. (unpublished)
%
\end{thebibliography}
\end{document}